# Creation of Single Chain of Nanoscale Skyrmion Bubbles with Record-high Temperature Stability in a Geometrically Confined Nanostripe

Zhipeng Hou[∥,†,‡], Qiang Zhang[∥,‡], Guizhou Xu[∥,§], Chen Gong[‡], Bei Ding[†], Yue Wang[†], Hang Li[†], Enke Liu[†], Feng Xu[§], Hongwei Zhang[†], Yuan Yao[†], Guangheng Wu[†], Xi-xiang Zhang*[‡], and Wenhong Wang*[†]

[†]Beijing National Laboratory for Condensed Matter Physics, Institute of Physics, Chinese Academy of Sciences, Beijing 100190, China

[‡]King Abdullah University of Science and Technology (KAUST), Physical Science and Engineering, Thuwal 23955-6900, Saudi Arabia

[§]School of Materials Science and Engineering, Nanjing University of Science and Technology, Nanjing 210094, China

[∥]These authors contributed equally to this work.

*Corresponding Author:
E-mail: wenhong.wang@iphy.ac.cn. Fax: +86-010-82649485
E-mail: xixiang.zhang@kaust.edu.sa. Fax: +966-12-8020221

**ABSTRACT:** Nanoscale topologically non-trivial spin textures, such as magnetic skyrmions, have been identified as promising candidates for the transport and storage of information for spintronic applications, notably magnetic racetrack memory devices. The design and realization of single skyrmion chain at room temperature (RT) and above in the low-dimensional nanostructures are of great importance for future practical applications. Here, we report the creation of a single skyrmion bubble chain in a geometrically confined $Fe_3Sn_2$ nanostripe with a width comparable to the featured size of a skyrmion bubble. Systematic investigations on the thermal stability have revealed that the single chain of skyrmion bubbles can keep stable at temperatures varying from RT up to a record-high temperature of 630 K. This extreme stability can be ascribed to the weak temperature-dependent magnetic anisotropy and the formation of edge states at the boundaries of the nanostripes. The realization of the highly stable skyrmion bubble chain in a geometrically confined nanostructure is a very important step towards the application of skyrmion-based spintronic devices.

Magnetic skyrmions are topologically protected vortex-like spin textures that were theoretically predicted two decades ago.[1] Recently, they have been experimentally observed in a small class of non-centrosymmetric magnets with a helical magnetic ground state caused by the Dzyaloshinskii-Moriya interaction (DMI).[2-8] Since the first experimental confirmation made by Mühlbauer *et al.*[2] with neutron scattering and subsequent confirmation of ultralow threshold for current-driven motion ($10^5 \sim 10^6 Am^{-2}$)[9] in the chiral magnet MnSi, magnetic skyrmions have been considered as promising carriers of information for spintronic applications,[10-16] particularly the racetrack memory devices.[11,17] However, the overall low thermal stability of magnetic skyrmions has been one of the major factors hindering their practical applications.[2,4-7] Recent experimental studies have demonstrated that their temperature window can be expanded significantly by means of zero-field cooling (ZFC),[18,19] field cooling (FC),[20-24] lowering the dimensionality,[3,25-31] electric field,[32] or pressure.[33] Remarkably, Du *et al.*[31] have realized a single skyrmion chain in a geometrically confined FeGe nanostripe over a wide temperature range from 100 K to 250 K, which is a significant progress towards the fabrication of skyrmion-based racetrack memory devices. However, for practical applications, the single skyrmion chain requires to be created and manipulated not only within a wide temperature range, but also at RT and even well above.

Alternatively, several groups have recently reported the observation of so-called magnetic skyrmion bubbles in the centrosymmetric magnets, such as Sc-doped hexagonal barium ferrite,[34] bilayered magnetite,[35] $Ni_2MnGa$,[36] MnNiGa,[37] $La_{1-x}Sr_xMnO_3$ (x=0.175),[38] and the frustrated magnet $Fe_3Sn_2$,[39] in which the skyrmion bubbles are stabilized by a competition between the magnetic dipole interaction and uniaxial anisotropy. Skyrmions and skyrmion bubbles are topologically equivalent to each other and possess some common topological properties, such as the topological Hall effect,[37,40] skyrmion Hall effect,[41,42] and ultra-low current density that controls their motion.[9,35] Most importantly, the materials with skyrmion bubbles usually exhibit a higher magnetic ordering temperature and the exchange interaction is generally stronger than the DMI.[43] These features make skyrmion bubbles can be stabilized, not only over a much wider temperature range but also at much higher temperatures. Moreover, recent theoretical simulations have shown that the formation and stability of individual skyrmion bubbles in the nanostructures of frustrated magnets are strongly affected by the edge states at the boundaries.[42] Therefore, experimentally create a single chain of skyrmion bubbles in the nanostructures and further investigate their temperature stability, are pivotal for both the applications of skyrmion-based racetrack memory and the fundamental insight to the dynamical mechanism of skyrmion bubbles.

Here, we report the design and experimental fabrication of nanostripes with well-defined geometries based on the frustrated magnet $Fe_3Sn_2$ (detailed in the Supporting information, SI, Note 1). Intriguingly, a single chain of skyrmion bubbles was created in the $Fe_3Sn_2$ nanostripe with a width comparable to the featured size of the bubbles. Further systematic investigations on the thermal stability showed that the single chain could keep stable at elevated temperatures, as high as 630 K. To elucidate the mechanisms behind the high stability, we investigated the morphology of the skyrmion bubbles and the evolution of their spin texture under different external magnetic fields, by using high-resolution Lorentz TEM. It was demonstrated that the record-high thermal stability could be attributed to the weak temperature-dependent magnetic anisotropy of $Fe_3Sn_2$ and the edge states resulting from the geometrically confined effects.

$Fe_3Sn_2$ is a rare example of the frustrated ferromagnetic compounds exhibiting a high Curie temperature $T_c$ up to 640 K (see Figure 1a). It undergoes a spin reorientation during which the easy axis rotates gradually from *c*-axis to *ab*-plane as the temperature decreases from 640 K to 100 K.[44-47] This compound has a layered rhombohedral structure with alternate stacking of the Sn layer and the Fe-Sn bilayers along the *c*-axis. The Fe atoms form bilayers of offset kagome networks with Sn atoms throughout the kagome layers, as well as between the kagome bilayers.[45] In this work, we synthesized the high-quality $Fe_3Sn_2$ single crystals by a Sn-flux method (Detailed information relating to the growth of the single crystals and their characterizations is given in the SI and Figure S1). Figure 1b shows the temperature-dependent saturation magnetization ($M_s$) of a $Fe_3Sn_2$ single crystal in the temperature range of 700-10 K under a magnetic field applied along the $c$-axis. It is observed that their values coincided with previous reports[46] and decreased monotonically as the temperature increased to $T_c$. Moreover, the overall behavior of $M_s$ (T) can be well described by the Bloch formula: $M_s(T)=M_s(0)(1-bT^{3/2})$, which is related to the localization of the magnetism.[48]

To investigate the influence of geometric confinement on the formation and stability of skyrmion bubbles, two $Fe_3Sn_2$ nanostripes with different widths were fabricated from a bulk single crystal by using focused ion beam (FIB) technique (Figure S2, SI). The scanning electron microscope (SEM) images demonstrate that both the nanostripes are flat and uniform (Figure S3, SI), suggesting a good quality. One nanostripe is designed to be 4000 nm in width, which is fifteen times larger than the diameter $D \approx 280$ nm of a magnetic skyrmion bubble reported in our previous work,[39] to represent a sample in which the skyrmion bubbles are not geometrically confined. The width of the other one is confined to be only 600 nm which is only approximately twice the value for $D$. It is expected that the formation of the skyrmion bubbles in the 600 nm nanostripe could be significantly influenced by the geometric confinement.

The upper panel of Figure 1c shows atypical SEM image of the 4000 nm nanostripe used in this study. To present its structure details more clearly, we have further shown a scanning transmission electron microscopy (STEM) image of the nanostripe in the left lower panel of Figure 1c. It should be noted here that the outer parts of the nanostripe are coated with amorphous carbon (black region) and platinum (grey region), which are applied to protect the $Fe_3Sn_2$ layer during the fabricating process and to reduce the interfacial Fresnel fringes in the Lorentz TEM image. Selected-area electron diffraction (SAED) was performed on the $Fe_3Sn_2$ layer to determine its orientation. The diffraction spots (the right lower panel of Figure 1c) showed a perfect six-fold symmetric characteristic, suggesting that the normal direction was along the [001]-orientation. In Figure 1d, we show the high-resolution scanning transmission electron microscopy image along the [001] direction. It is clear to see that Fe (blue particles) and Sn atoms (red particles) arranged into the kagome lattice in an alternate fashion (as shown in the upper-right inset), which agrees well the crystal structure of $Fe_3Sn_2$ and confirms the nanostripes are in a good quality.

Hereafter, we investigated the evolution of magnetic domains in the 4000 nm nanostripe at room temperature by using Lorentz TEM equipped with a spherical aberration corrector for imaging system (Titan G2 60–300, FEI) at an acceleration voltage of 300 kV. Figure 1e shows the under-focused Lorentz TEM image taken at a zero magnetic field. The sinusoidal variation of the Lorentz TEM contrast suggests that the helix period is about 430 nm (see the inset of Figure 1e). By gradually increasing the magnetic field along the [001] direction from zero, the magnetic stripe domains underwent a complex process and eventually transformed into skyrmion bubbles in a critical field of 300 mT (see Figure 1f). It is found that the transforming process, the morphology of the magnetic bubbles and even the value of the critical field (Figure S4, SI) are quite similar to that observed in the $Fe_3Sn_2$ thin plate.[38] Therefore, we can conclude that the 4000 nm nanostripe can be considered as a comparison sample in which the geometric confinement show little influence on the formation of magnetic skyrmion bubbles.

Since the $T_c$ of $Fe_3Sn_2$ is as high as 640 K, it is natural to expect that the skyrmion bubbles can survive up to a higher temperature. To obtain Lorentz TEM images in the high-temperature region, a double-tilt heating holder (Model 652, Gatan Inc.) with a smart set hot stage controller (Model 901, Gatan Inc.) were used to rise the specimen temperature from 300 K to 670 K. In Figure 2a-f, we show the temperature-dependent under-focused Lorentz TEM images of skyrmion bubbles at temperatures ranging from 300 K to 573 K under their corresponding critical magnetic fields (all the magnetic stripes transform into skyrmion bubbles under the magnetic field). The detailed transformation processes of the magnetic stripe domains are presented in Figure S5 in the Supporting information. One can notice that the critical field decreased correspondingly with the increase of temperature. More intriguingly, both the diameter and density of the magnetic bubbles remain almost unchanged, even at a temperature as high as 523 K, indicating a high thermal stability of the skyrmion bubbles in the $Fe_3Sn_2$ nanostripe. However, when the temperature increased above 553 K, the density of the skyrmion bubbles decreased substantially, which suggests that the thermal fluctuations start to make the spin structrue of skyrmion bubbles unstable. At 573 K, when we increased the applied magnetic field, the magnetic stripes vanished directly without going through the phase of skyrmion bubble (see Figure 2f).

In the case of the 600 nm $Fe_3Sn_2$ nanostripe, the evolution process of the magnetic domains involved a gradual transformation of the magnetic stripes into individual skyrmion bubbles when the applied magnetic field increased. Typical Lorentz TEM images of this transformation are given in Figure 3a-e. By using the transport-of-intensity equation (TIE) analysis, we present the corresponding spin textures of the magnetic domains enclosed by white boxes in Figure 3a, c, and d in their corresponding inset, respectively. At zero magnetic field, we found that a series of nanosized magnetic stripe domains arranged along the long axis of the nanostripe with an average periodicity of 360 nm, which is slightly smaller than that of 430 nm in the 4000 nm nanostripe. As the magnetic field increased, the magnetic stripes first became narrow, and then swirled into pocket-shape domains. When the magnetic field increased above 120 mT, the pocket-shape domains became skyrmion bubbles entirely. Once formed, the bubbles tended to move towards the interior (due to the repulsion with the edge spins) and assembled to form a single chain along the long axis of the nanostripe. By analyzing the spin distributions of the skyrmion bubbles in the 600 nm $Fe_3Sn_2$ nanostripe (see the inset of Figure 3d), they were found to converge, and therefore we proposed their corresponding topological number *N* to be 1. Those experimental results can be reproduced by a theoretical simulation of the evolution of the skyrmion bubbles with increasing the perpendicular field, as shown in Figure 3g-j, where the size of skyrmion bubbles are revealed to be gradually decrease with increasing fields. In addition, some of the skyrmion bubbles display elliptical shapes which are slightly distort from the circular ones. According to recent Monte Carlo simulations and Lorentz TEM experiments,[36,49,50] it is supposed that the distortion results from a small in-plane component of the oblique magnetic field. We have further confirmed this assumption by carrying out the angle-dependent Lorentz experiments (Figure S6, SI). It is clearly demonstrated that the skyrmion bubbles gradually become elliptically distorted by titling the sample. Notably, by careful high-resolution TEM analysis (Figure S7, SI), a small strain was revealed in the fabricated $Fe_3Sn_2$ nanostripe. We propose that it may also induce the deformations of skyrmion bubbles, as is demonstrated in the chiral magnet of FeGe.[51]

We have further checked the thermal stability of the single skyrmion bubble chain over a wide temperature range from 300K to $T_c$ (640 K). Figure 4a-d shows the under-focused Lorentz TEM images of magnetic bubbles at different temperatures under their corresponding critical magnetic field (i.e. the applied field needed for the formation of skyrmion bubbles). Intriguingly, we found that the single skyrmion bubble chain could keep stable even up to an extremely high temperature of 630 K. In comparison with the critical temperature observed in the 4000 nm $Fe_3Sn_2$ nanostripe, the one in the 600 nm nanostripe is significantly enhanced by approximately 100 K. Our results thus strongly suggest that the width confinement is beneficial for the thermal stability of skyrmion bubbles. To present the high thermal stability in a more quantitative manner, as shown in the upper plane of Figure 4e, we described the ellipticity of the skyrmion bubble defining two parameters of the major semi-axis *a* and minor semi-axis *b*. By plotting the values of *a* and *b* as a function of temperature in lower plane Figure 4e, we found that both values were temperature-independent, indicating that the size of the skyrmion bubbles remained almost a constant over the wide temperature range from 300 K to 630 K. Moreover, the average distance *d* between two neighboring skyrmion bubbles also kept unchanged within this wide temperature range, further confirming the high stability of the single skyrmion bubble chain. The realization of the highly stable single skyrmion bubble chain in the geometrically confined $Fe_3Sn_2$ nanostripes can be attributed to: (i) the weak temperature-dependent magnetic anisotropy $K_u$ of the $Fe_3Sn_2$ crystal that the value of $K_u$ remains rather stable as temperature increases from 300 K to 630 K (Figure S8, SI); (ii) the formation of edge states at the boundaries of the nanostripes. For the frustrated magnets, edge states originate from a strong easy-plane surface anisotropy due to the lower symmetry of magnetic irons at the boundaries.[42] In the formation process of skyrmion bubbles, it can act as the initial nucleation centers, as is the case in the highly geometric-confined FeGe nanostripe.[31,52] This feature not only helps the skyrmion bubbles form at a relatively low field (Figure S9, SI), but also enhance their temperature stability in the 600 nm nanostripe. We should note that, the critical field to form skyrmion bubbles in the $Fe_3Sn_2$ nanostripe is larger than that in the polar magnet $GaV_4S_8$ with strong anisotropy,[7] magnetic films with interfacial DMI interaction,[16,52] chiral magnets $Fe_{0.5}Co_{0.5}Si$,[4] and $Cu_2OSeO_3$.[6] However, very recently, both the zero-field stable skyrmions and skrymion bubbles have been realized via a simple field cooling manipulation in the Co-Mn-Zn[21,22] and MnNiGa,[23] respectively. It is hoped that we can realize the single skyrmion bubble chain without any support of external magnetic fields in the $Fe_3Sn_2$ nanostripes in our future work.

In conclusion, we have successfully created a highly stable single chain of skyrmion bubbles in a geometrically confined $Fe_3Sn_2$ nanostripe. In contrast to other centrosymmetric or non-centrosymmetric materials, the highest stable temperature of the skyrmion bubbles in the 600 nm $Fe_3Sn_2$ nanostripe is record-high, as shown in Figure 5. This extreme stability can be attributed to the weak temperature-dependent magnetic anisotropy and the formation of edge states at the boundaries of the nanostripes. The realization of a single chain of skyrmion bubble in the $Fe_3Sn_2$ nanostripe over an extremely wide temperature range well above room temperature is a very important step towards the fabrication of skyrmion-based racetrack memory devices.

**Acknowledgment**

This work was supported by the National Key R&D Program of China (Grant No. 2017FA0303202), National Natural Science Foundation of China (Grant Nos. 11604148, 1561145003, 11574374), King Abdullah University of Science and Technology (KAUST) Office of Sponsored Research (OSR) under Award No: CRF-2015-2549-CRG4, and Strategic Priority Research Program B of the Chinese Academy of Sciences under the grant No. XDB07010300.

**Supporting Information**

● S1: Characterization of the $Fe_3Sn_2$ single crystals.

● S2: Fabrication process of the $Fe_3Sn_2$ nanostripes.

● S3: The thickness at different positions on the nanostripe.

● S4: Under-focused LTEM images under different out-of-plane magnetic fields in the 4000nm $Fe_3Sn_2$ nanostripe at 300K.

● S5: Under-focused LTEM images under different magnetic fields in the 4000nm $Fe_3Sn_2$ nanostripe measured at various temperatures between 373K and 573K.

● S6: The angle-dependent LTEM images for the 600 nm nanostripe.

● S7: High resolution TEM measurements.

● S7: Magnetic measurements of bulk $Fe_3Sn_2$ single crystal.

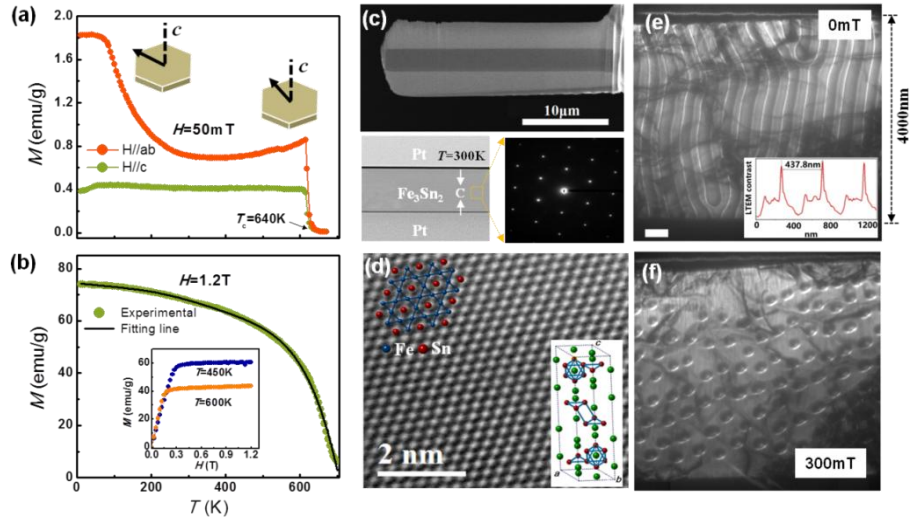

**Figure 1.** Basic characteristics of the bulk $Fe_3Sn_2$ single crystal and 4000 nm nanostripe (a) Temperature dependence of low-field (0.05 T) magnetization for single crystal of $Fe_3Sn_2$. The Curie temperature is determined to be approximately 640 K. (b) Temperature dependence of saturated magnetization curve with magnetic field applied along *c*-axis. The same data are obtained when the field is applied along *ab*-plane. The black solid line is the fitting of the $M_s(T)$ to the Bloch equation: $M_s(T)=M_s(0)(1-bT^{3/2})$. (c) A typical scanning electron microscopy (STEM) image of the 4000 nm wide nanostripe and selected-area diffraction pattern are shown in the left and right lower panels, respectively. (d) High-resolution scanning transmission electron microscopy image taken on the [001] axis. The left inset shows the alternate arrangement of Fe (blue particles) and Sn atoms (red particles) into the kagome lattice, whereas the right inset displays the crystal structure of the $Fe_3Sn_2$. (e) Room-temperature under-focused LTEM images taken at a zero field showing the existence of a helical magnetic structure. The inset depicts the sinusoidal change in the LTEM contrast with distance as marked by the white line in the main panel. (f) Room-temperature under-focused LTEM images taken in a field of 300 mT showing the existence of skyrmion bubbles.

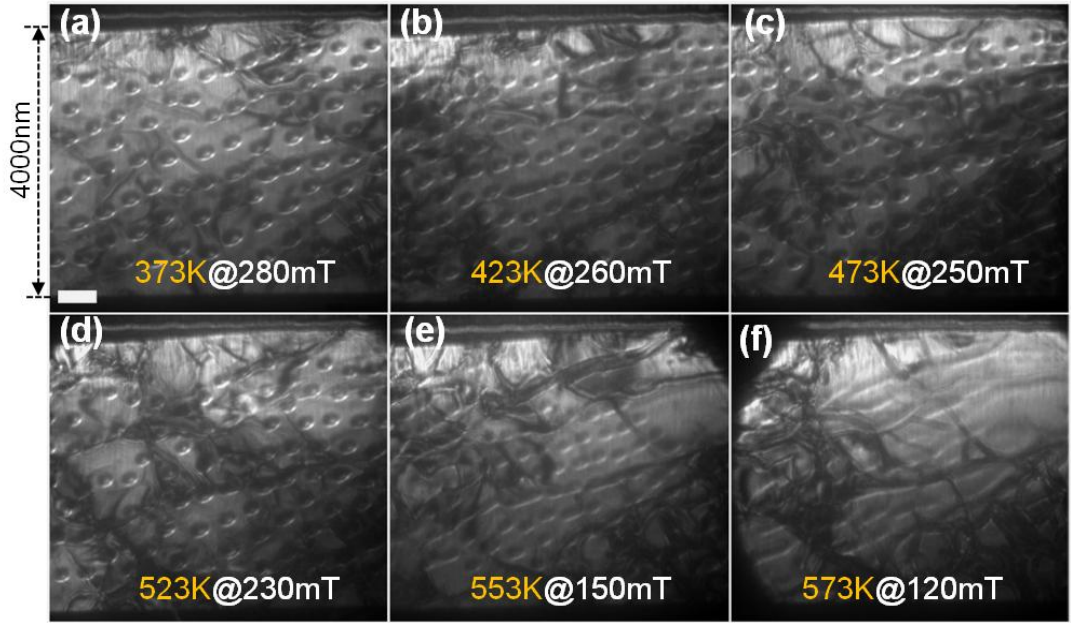

**Figure 2.** Temperature dependence of magnetic bubbles in the 4000 nm nanostripe. (a-f)Under-focused LTEM images of the magnetic bubbles in the 4000 nm nanostripe taken at temperatures ranging from 300 K to 573 K under their corresponding critical magnetic field. All the photos are taken in the same region of the nanostripe. Notably, when the temperature rises above 473 K, some dendrite impurities are brought by thermal fluctuation and can adhere to the surface of the sample. The scale bar is 500nm in all the images.

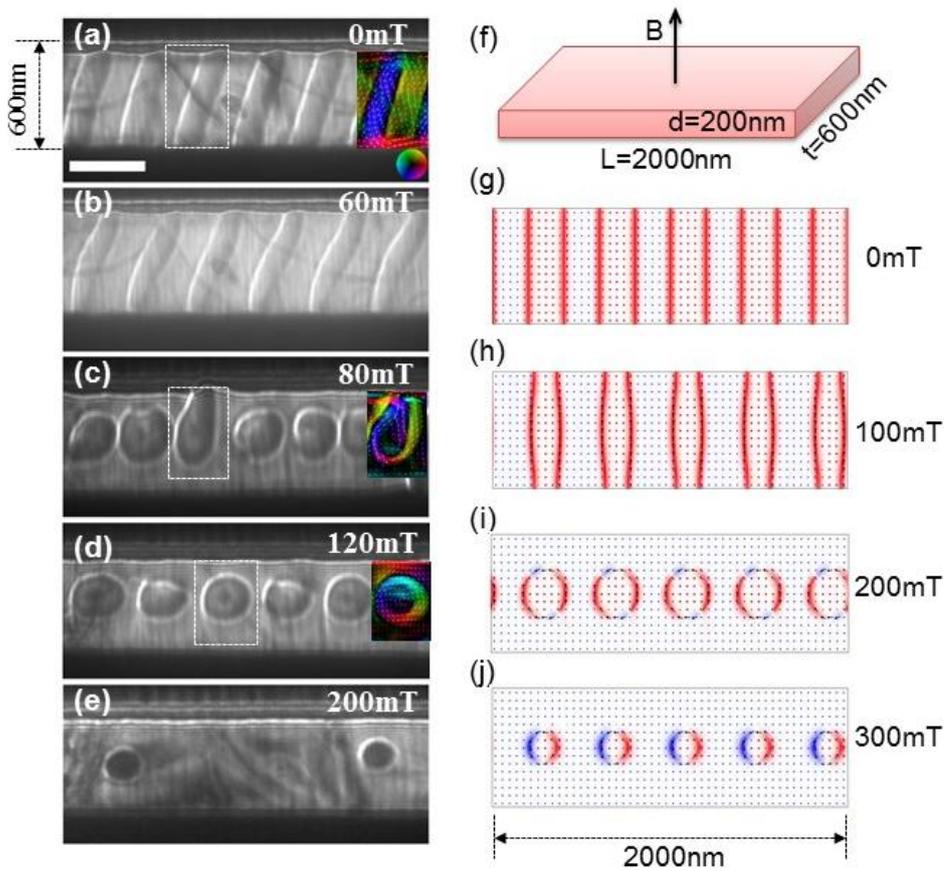

**Figure 3.** The evolutionary process of magnetic domains under different out-of-plane magnetic fields in a 600 nm nanostripe at 300 K. (a)-(e) The under-focused LTEM images under different out-of-plane magnetic fields. The domain enclosed by white boxes present a one-to-one correspondence in the stripe-bubble transformation. Inset: the corresponding magnetization textures obtained from the TIE analysis for the domains enclosed by white boxes. The scale bar is 500nm. (f)-(j) Micromagnetic simulation of the evolution of the single chain of skyrmion bubbles as a function of the magnetic field.

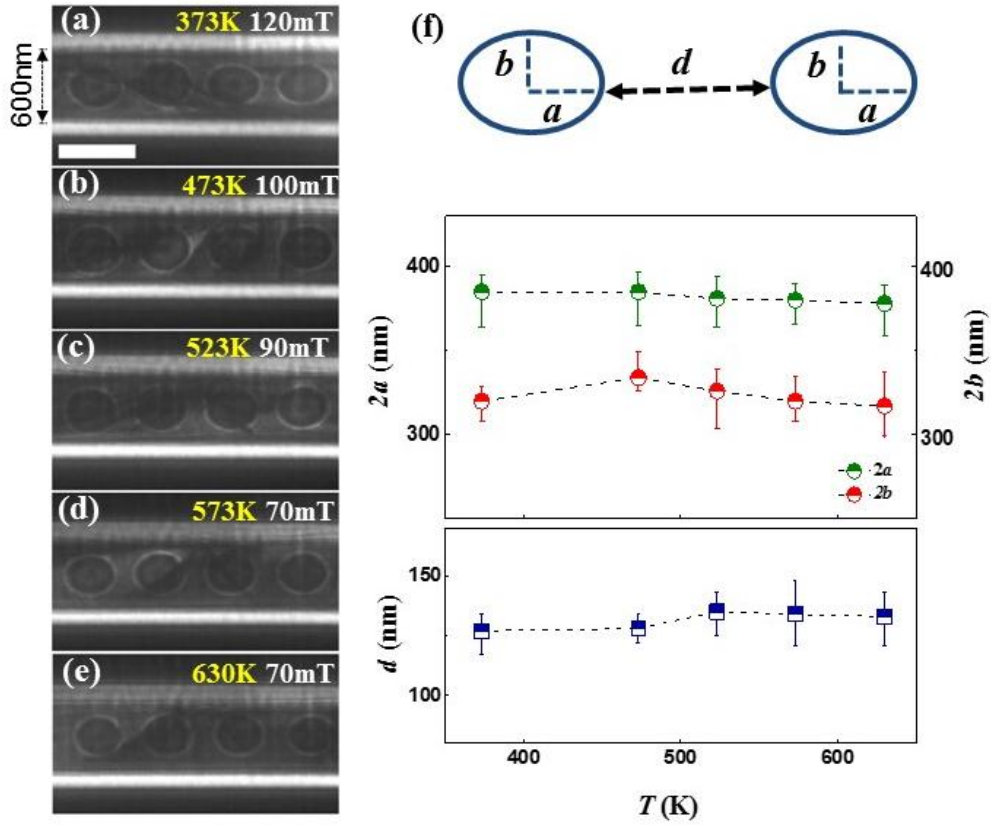

**Figure 4.** Temperature stability of the skyrmionic bubbles in a 600 nm nanostripe at temperatures ranging from 373 K to 630 K. (a)-(d) Under-focused LTEM images of skyrmionic bubbles in the 600 nm nanostripe taken at temperatures ranging from 300 K to 630 K under their corresponding critical magnetic field. All the photos are taken in the same region of the nanostripe. The scale bar is 500nm. (e) Upper plane: Schematic diagram illustrating the morphology of skyrmion bubbles and the distance between them. The shape of the bubble is assumed to be elliptic shape. The half-major axis, half-minor axis, and distance between the bubbles are represented by $a$, $b$, and $d$, respectively. Lower plane: Temperature-dependent values of $2a$, $2b$, and $d$. The half-filled circles represent the average value; the error bars are added based on results summarized from (a)-(d).

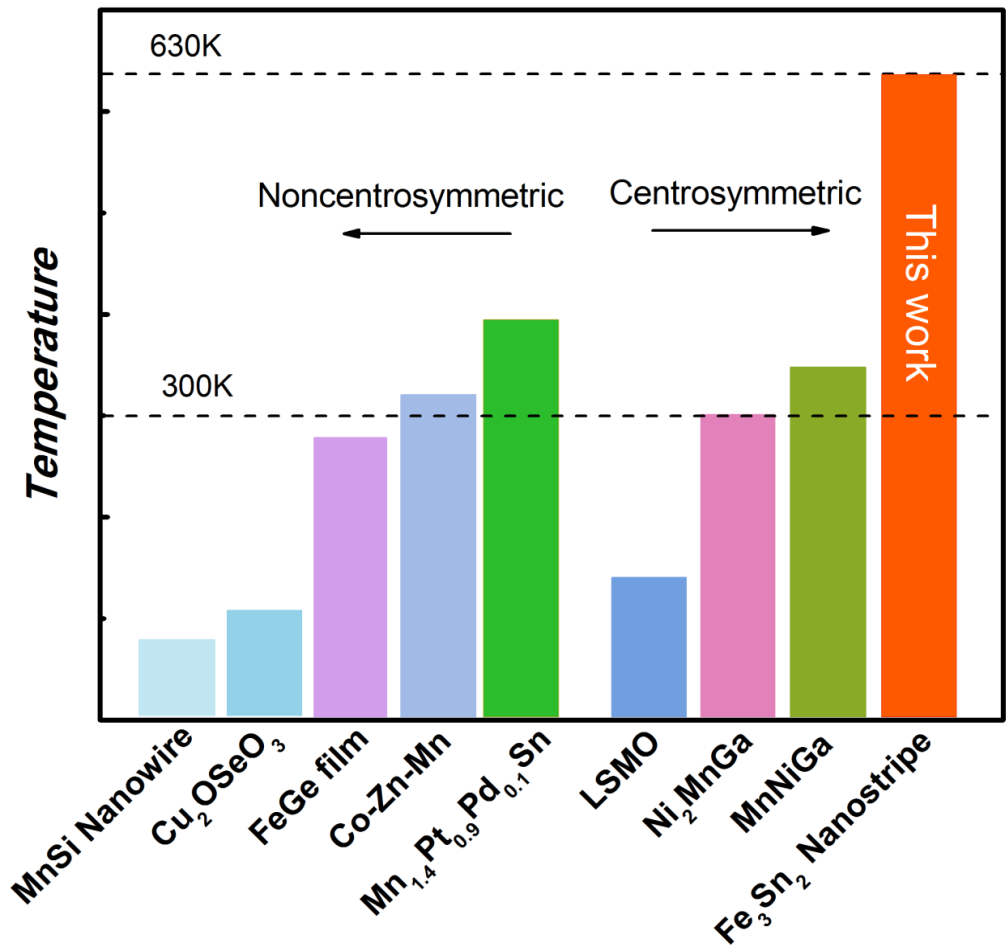

**Figure 5.** Comparison of the highest stable temperature of the skyrmions and skyrmion bubbles in non-centrosymmetric and centrosymmetric materials, respectively. The data are taken from the published studies, including Refs. (3), (5), (29), (31), (35-37) and (54) based on the *in-situ* Lorentz TEM observations.

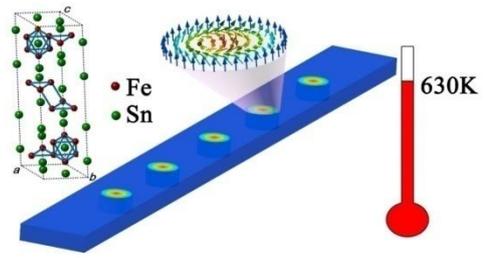

**for TOC only**